\documentclass[singlecolumn,showpacs,preprintnumbers,amsmath,amssymb]{revtex4}
\usepackage{graphicx}
\usepackage{dcolumn}
\usepackage{bm}

\begin{document}

\title{Analytical parametrization of fusion barriers using proximity potentials}

\author{Ishwar Dutt }
\author{Rajeev K. Puri}%
\email{rkpuri@pu.ac.in;drrkpuri@gmail.com} \affiliation{Department
of Physics, Panjab University, Chandigarh 160014, India}


\date{\today}

\begin{abstract}

Using the three versions of proximity potentials, namely proximity
1977, proximity 1988, and proximity 2000, we present a pocket
formula for fusion barrier heights and positions. This was
achieved by analyzing as many as 400 reactions with mass between
15 and 296. Our parametrized formula can reproduced the exact
barrier heights and positions  within an accuracy of $\pm1\%$. A
comparison with the experimental data is also in good agreement.
\end{abstract}

\pacs{24.10.-i, 25.70.Jj, 25.70.-z.}

\maketitle


\section{\label{intro}Introduction}
 In the low energy heavy-ion collisions, fusion of colliding nuclei
 and related phenomena has always been of central interest\cite{rkp1}.
 Depending upon the incident energy of the projectile as well as angular momentum and
 impact parameter, the collision of nuclei can lead to several
 interesting phenomena such as  incomplete fusion~\cite{rkp1},
 multifragmentation~\cite{rkp2,rkp3}, subthreshold particle production~\cite{rkp4}, nuclear flow~\cite{rkp5} as well as formation
 of the superheavy elements~\cite{rkp6}. Since fusion is a low density
 phenomenon, several mean field models~\cite{rkp1,rkp6,blocki77,wr94,ms2000,wang06,deni02} have been developed in the recent past at
 microscopic/macroscopic level and have  been robust against the vast
 experimental data~\cite{ms2000,wang06,expt} that range from symmetric to highly asymmetric colliding nuclei. The study of mass
 dependence has always guided the validity of various models irrespective of the energy range. The essential idea of developing a model
is to understand the physical mechanism behind a process or
phenomenon. Extension of the physics is also reported toward
isospin degree of freedom. At the same time, accumulation of huge
experimental data~\cite{ms2000,wang06,expt} (that include all
kinds of masses and asymmetry of colliding nuclei) puts stringent
test for any theoretical model.
\par
As fusion process occurs at the surface of colliding nuclei, any
difference occurring in the interior part of the potential does
not make any difference toward the fusion. One always tries to
parametrize the potential in terms of some known quantities such
as the masses and charges of colliding
nuclei~\cite{rkp1,deni02,bass73,ngo75}. At intermediate energies,
several forms of density dependent potentials are also
available~\cite{rkp2,rkp3,rkp4,rkp5}. Generally, the benchmark is
to parameterized the outcome in proximity fashion~\cite{blocki77}.
By adding the Coulomb potential to the parameterized form of the
nuclear ion-ion potential, one obtains total ion-ion potential and
ultimately, the fusion barriers and cross sections.
\par
Alternatively, one calculates the barrier heights as well as
positions of large number of reactions and then tries to
parametrize these in terms of some known quantities like the
charges and masses of the colliding nuclei~\cite{rkp1,skg82}.
Recently, even neutron excess dependence has also been
incorporated in some attempts~\cite{ng04}. Similarly, an
analytical expression  to determine the barrier heights and
positions are also presented in Ref.~\cite{rm01}. The cost of such
attempts was in the form of more complicated parametrized form.
The utility of such direct parametrization is that one can use
these pocket formula to find out the fusion barriers
instantaneously.
\par
As is evident from the literature, several modifications over the
original proximity potential have also been suggested in the
recent years~\cite{wr94,ms2000}. We shall here attempt to present
a direct parametrization of the fusion barrier positions as well
as heights using different proximity potentials. This attempt will
introduce great simplification in obtaining the fusion barrier
positions and heights. Section \ref{model} describes the models in
brief, Sec. \ref{result} depicts the results, and a summary is
presented in Sec. \ref{summary}.
\section{\label{model} The Model}
All proximity potentials are based on the proximity force theorem.
According to which, \emph{``the force between two gently curved
surfaces in close proximity is proportional to the interaction
potential per unit area between the two flat surfaces''}. The
nuclear part of the interaction potential in different proximity
potentials is described as a product of geometrical factor
representing the mean curvature of the interacting surfaces and an
universal function depending on the separation distance.
\subsection{$\rm Proximity~1977~(Prox~77)$}
 According to the original version of proximity~\cite{blocki77}, the interaction potential $ V_{N}(r)$ between two
 surfaces can be written as
 \begin{equation}
V_{N}^{Prox~77}(r)= 4\pi \gamma b \overline{R} \Phi \left( \frac{{
r}-C_{1} -C_{2}}{b}\right) {~\rm MeV},\label{eq:1}
\end{equation}
 where the surface energy coefficient $\gamma$ taken from the Lysekil
 mass formula $( ~\rm in
 ~MeV/fm^{2})$ is written as
 \begin{equation}
\gamma = \gamma_{0}\left[1-k_{s} I^{2} \right],
 \label{eq:2}
\end{equation}
with $ I=\left(\frac{N-Z}{A}\right)$; $N$, $Z$, and $A$ refer to
the neutron, proton and total mass of two interacting nuclei.
Though the proximity potential Prox 77, in principle, is for
zero-neutron excess, the factor $ \gamma$ takes care of some
neutron excess content. In the above formula, $ \gamma_{0}$ is the
surface energy constant and $k_{s}$ is the surface-asymmetry
constant. Both constants were first parametrized by Myers and
\'Swi\c{a}tecki~\cite{ms66} by fitting the experimental binding
energies. The first set of these constants yielded values
$\gamma_{0}$ and $k_{s}=1.01734 ~\rm
 ~MeV/fm^{2}$ and 1.79, respectively. Later on, these
values were  revised to ${~\rm \gamma_{0}}$ = 0.9517 $~\rm
 MeV/fm^{2}$ and
$ k_{s}=1.7826$~\cite{ms67}. Interestingly, most of the modified
proximity type potentials use different values of the parameter
$\gamma$~\cite{wr94,ms2000}. The mean curvature radius, $
\overline{R}$ in Eq. (\ref{eq:1}) has the form
\begin{equation}
\overline{R} = \frac{C_{1}C_{2}}{C_{1}+ C_{2}}, \label{eq:3}
\end{equation}
quite similar to the one used for reduced mass. Here
\begin{equation}
C_{i}= R_{i}\left[1-\left(\frac{b}{R_{i}} \right)^{2}+\cdots
\cdots \right], \label{eq:4}
\end{equation}
${\rm R_{i}}$, the effective sharp radius, reads as
\begin{equation}
R_{i}= 1.28A^{1/3}_{i}- 0.76+0.8A^{-1/3}_{i} {~\rm fm}~~(i=1,2).
\label{eq:5}
\end{equation}
 The universal function $\Phi \left(\xi
\right)$ was parametrized with the following form:
\begin{equation}
\Phi \left(\xi \right)= \left\{
\begin{array}{l}
-\frac{1}{2} \left(\xi- 2.54 \right)^{2}-0.0852\left(\xi- 2.54
\right)^{3},\\
~~~~~~~~~~~~~~~~~~~~~~~ \mbox{ for $\xi \leq 1.2511 $ },    \\
-3.437\exp \left(-\frac{\xi}{0.75} \right),\\
~~~~~~~~~~~~~~~~~~~~~~~ \mbox{ for $\xi \geq  1.2511 $ },
\end{array}
\right. \label{eq:6}
\end{equation}
with $\xi$ = $(r - C_{1} - C_{2}$)/$b$. The width $b$ has been
evaluated close to unity. Using the above form, one can calculate
the nuclear part of the interaction potential ${ V_{N}(r)}$. This
model is referred as Prox 77 and corresponding potential as $
V_{N}^{Prox~77}(r)$.
\subsection{$\rm Proximity~1988~(Prox~88)$}
Later on, using the more refined mass  formula of M\"oller and
Nix~\cite{mn81}, the value of coefficients $\gamma_{0}$ and $
k_{s}$ were modified yielding their values =1.2496 $\rm
MeV/fm^{2}$ and 2.3, respectively. Reisdorf \cite{wr94} labeled
this modified version
 as `Proximity 1988'. Note that this set of
 coefficients give stronger attraction  compared to the above sets. Even a more recent compilation
 by M\"oller and  Nix \cite{mn95} yields similar values. We
marked this potential as Prox 88.
\subsection{$\rm Proximity~2000~(Prox~00)$}
Recently, Myers and \'Swi\c{a}tecki~\cite{ms2000} modified
Eq.~(\ref{eq:1}) by using up-to-date knowledge of nuclear radii
and surface tension coefficients using their droplet model
concept. The prime aim behind this attempt was to remove
discrepancy of the order of $4\%$ reported between the results of
Prox 77 and experimental data~\cite{ms2000}.
 Using the droplet model
\cite{ms80}, matter radius $C_{i}$ was calculated as
\begin{equation}
C_{i}= c_{i}+ \frac{N_{i}}{A_{i}}t_{i}    ~~~~(i=1,2),
\label{eq:7}
\end{equation}
where $c_{i}$ denotes the half-density radii  of the charge
distribution and $t_{i}$ is the neutron skin of the nucleus. To
calculate $c_{i}$, these authors~\cite{ms2000} used  two-parameter
Fermi function values given in Ref. \cite{dv87} and remaining
cases were handled with the help of parametrization of charge
distribution described below. The nuclear charge radius (denoted
as $R_{00}$ in Ref.~\cite{bn94}), is given by the relation:
\begin{equation}
R_{00i}=\sqrt{\frac{5}{3}}\left<r^{2}\right>^{1/2} \nonumber
\end{equation}
\begin{eqnarray}
 \nonumber \\
=1.240A_{i}^{1/3} \left\{1+\frac{1.646}{A_{i}}-0.191
\left(\frac{A_{i}-2Z_{i}}{A_{i}}\right)\right\} {~\rm fm}~~ \nonumber \\
(i=1,2), \label{eq:8}
\end{eqnarray}
where $<r^{2}>$ represents the mean square nuclear charge radius.
According to Ref.~\cite{bn94}, Eq.~(\ref{eq:8}) was valid for the
even-even nuclei with $8\leq Z < 38$ only. For nuclei with $Z\geq
38$, the above equation was modified by Pomorski \emph{et
al}.~\cite{bn94} as
\begin{equation}
R_{00i}=  1.256A_{i}^{1/3}
\left\{1-0.202\left(\frac{A_{i}-2Z_{i}}{A_{i}}\right)\right\}
{~\rm fm}. \label{eq:9}
\end{equation}
These expressions give good estimate of  the measured mean square
nuclear charge radius $<r^{2}>$. In the present model, authors
used only  Eq.~(\ref{eq:8}). The half-density radius, $c_{i}$ was
obtained from the relation:
\begin{equation}
c_{i}= R_{00i}\left(1-
\frac{7}{2}\frac{b^{2}}{R_{00i}^{2}}-\frac{49}{8}\frac{b^{4}}{R_{00i}^{4}}+\cdots
 \right) ~~~~~~~(i=1,2). \label{eq:10}
\end{equation}
 Using the droplet
model~\cite{ms80}, neutron skin $t_{i}$ reads as
\begin{equation}
t_{i}= \frac{3}{2}r_{0}\left[\frac{JI_{i}-
\frac{1}{12}c_{1}Z_{i}A^{-1/3}_{i}}{Q+ \frac{9}{4}JA^{-1/3}_{i}}
\right]     (i=1,2). \label{eq:11}
\end{equation}
Here $r_{0}$ is $1.14$ fm, the value of nuclear symmetric energy
coefficient $J=32.65$ MeV and $c_{1}= 3 e^{2}/5 r_{0}=0.757895$
MeV. The neutron skin stiffness coefficient $Q$ was taken to be
35.4 MeV. The nuclear surface energy coefficient $\gamma$ in terms
of neutron skin was given as;
\begin{equation}
\gamma = \frac{1}{4\pi r^{2}_{0}}\left[18.63  {\rm  (MeV)}
-Q\frac{\left(t^{2}_{1} + t^{2}_{2}\right)}{2r^{2}_{0}} \right],
\label{eq:12}
\end{equation}
where $t_{1}$ and $t_{2}$ were calculated using Eq.~(\ref{eq:11}).
 The universal function $\Phi
(\xi)$ is reported as
\begin{equation}
\Phi \left(\xi \right)=\left \{
\begin{array}{ll}
-0.1353+ \sum\limits_{n=0}^{5}\left[c_{n}/\left(n+1\right)\right]
\left(2.5 - \xi \right)^{n+1},\\
~~~~~~~~~~~~~~~~~~~~~~~~~~~ \mbox{ for \quad $0 < \xi \leq   2.5 $},\\
-0.09551\exp \left[\left( 2.75 - \xi\right)/0.7176 \right], \\
~~~~~~~~~~~~~~~~~~~~~~~~~~~\mbox{ for $\quad\xi \geq 2.5$}.
\end{array}
\right. \label{eq:13}
\end{equation}
The values of different constants $c_{n}$ were: $c_{0}=-0.1886$,
$c_{1}=-0.2628$, $c_{2}=-0.15216$, $c_{3}=-0.04562$,
$c_{4}=0.069136$, and $c_{5}=-0.011454$. For $\xi
> 2.74$, the above exponential expression is the exact representation of
the Thomas-Fermi extension of the proximity potential. This
potential is marked  as Prox 00.

\section{\label{result}Results and Discussion}
 As a first step, we calculated the nuclear part of the ion-ion
 potential using Prox 77, Prox 88, and Prox 00 potentials
 and then by adding the Coulomb potential~(= $\frac{Z_{1}Z_{2}e^{2}}{r}$), total ion-ion potential $V_{T}(r)$ for spherical colliding pair is
 obtained. The fusion barrier is then extracted using conditions
 \begin{equation}
\frac{dV_T(r)}{dr}|_{r=R_{B}} = 0,~~ {\rm{and}} ~~
\frac{d^{2}V_T(r)}{dr^{2}}|_{r=R_{B}} \leq 0. \label{eq:14}
\end{equation}
 The height of the barrier and position is marked, respectively, as $V_{B}$ and $R_{B}$. For the present analysis, all kind
 of the reactions involving symmetric $(N=Z,~ A_{1}=A_{2})$ as well as asymmetric
 $(N\neq Z,~ A_{1}\neq A_{2})$ nuclei are considered. In all, 400 reactions covering almost whole of the periodic table
 are taken into account. All nuclei considered here are assumed  to be  spherical in nature, however, deformation as well as orientation of the nuclei
 also affect the fusion barriers~\cite{deni07}. The lightest reaction considered here is $^{6}Li+^{9}Be$ whereas the heaviest one is $^{48}Ca+^{248}Cm$.
 As  reported in Ref.~\cite{ms2000}, proximity Prox 77 overestimate experimental data by $4\%$.  It
 was reported to be better  for newer versions.

\par
 Once fusion barrier heights and positions were
 calculated, a search was made for their parametrization.
Since it is evident that barrier positions depend on the size of
 the colliding systems, the best way is to parametrize them in terms
 of the
 radius dependence i.e. in terms of  $A^{1/3}$. In the literature,
 several  attempts exist that parametrize $R_{B}$ directly either as $A^{'} + B^{'} (A_{1}^{1/3}+A_{2}^{1/3})$~\cite{anjos02,broglia81,kovar79,cw76}
 or as
 $r_{B}~(=\frac{R_{B}}{A_{1}^{1/3}+A_{2}^{1/3}}$)~\cite{cngo75,ngo80}. We have also
 tried  similar fits. Unfortunately, the scattering around the
 mean curve  was quite significant in both the cases, therefore, we discard this kind of parametrizations. Alternatively, we plotted
the reduced fusion barrier positions $s_{B}=R_{B}-C_{1}-C_{2}$, as
a function of $\frac{Z_1 Z_2}{A_1^{1/3}+A_2^{1/3}}$ for all three
versions of proximity potentials~(see Fig. 1).
 Very encouragingly, the reduced barrier positions
$s_{B}$ of all the reactions fall
 on  the mean curve that can  be parametrized in terms of exponential
 function. We noted that the scattering around the mean positions is very small. Due to the
 weak Coulomb force in lighter colliding nuclei, lesser attractive potential is needed to
 counterbalance it. As a result, separation distance increases in lighter colliding nuclei.  As we go to
 heavier nuclei, stronger Coulomb contribution demands more and more
 penetration, therefore, decreasing the value of $s_{B}$. In other words, the fusion in
 lighter nuclei occurs at the outer region compared to the heavier nuclei
 where $s_{B}$ is much smaller.
 \par
  If we compare (a) and (b) parts of the Fig. 1, we notice that $s_{B}$, the
  separation distance between nuclei is slightly more in Prox 88
  compared to Prox 77. This is due to the fact that Prox 88 has stronger surface
  energy coefficient $\gamma$ [see Eq. (\ref{eq:2}) with $\gamma_{0}$ = $1.2496 ~\rm
MeV/fm^{2}$ and $ k_{s}=2.3$ respectively]. This
  results in more attractive nuclear potential compared to
  Prox 77 and therefore, counterbalancing happens at larger distances. From the figure,
  it is also evident that latest proximity potential has shallow nuclear potential compared to the other two versions.
  All three proximity potentials follow similar mass/ charge dependence and can be parametrized in terms of following
  function:
  \begin{equation}
{s_B^{par}} = \alpha\exp\left[-\beta
\left(x-2\right)^{1/4}\right]. \label{eq:15}
\end{equation}
Here, $x$ = $\frac{Z_1 Z_2}{A_1^{1/3}+A_2^{1/3}}$ and $\alpha$,
$\beta$ are the constants whose values depend on the model one is
using. The values of $\alpha$, are 5.184 19, 5.374 57, and 5.087
58, whereas the values of $\beta$ are 0.339 79, 0.313 26, and
0.295 18 for Prox 77, Prox 88, and Prox 00, potentials,
respectively. The analytical parametrized fusion barrier positions
therefore, read as
\begin{equation}
{R_B^{par}} = {s_B^{par}} + C_1 + C_2. \label{eq:16}
\end{equation}
 The quality of our parametrized fusion positions can be judged
by analyzing the percentage deviation defined as
\begin{equation}
\Delta R_{B}~(\%) = \frac{R_{B}^{par}-
R_{B}^{exact}}{R_{B}^{exact}}\times 100. \label{eq:17}
\end{equation}
We plot in Fig. 2, the percentage deviation $\Delta R_{B}~(\%)$ as
a function of the product of charges $Z_{1}Z_{2}$. Very
encouragingly, we see that  in all three cases, our analytical
parametrized form gives very good results within $\pm1\%$ of  the
actual exact barriers positions. The average deviations calculated
over 400 reactions are -0.01\%, -0.02\%, and 0\% for Prox 77, Prox
88, and Prox 00, respectively. This is very encouraging since it
is for the first time that such accurate parametrization has been
obtained. Note that our parametrizations depend on the charges and
masses of the colliding nuclei only. This definitely introduces
great simplification in the calculation of fusion barrier
positions within proximity concept.

\par
In Fig. 3, we parametrize the fusion barrier heights $V_{B}$ as a
function of $\frac{1.44
Z_1Z_2}{R_B^{par}}(1-\frac{0.75}{R_B^{par}})$, similar to the one
reported in Refs.~\cite{ng04,broglia81}. The first part is the
Coulomb contribution whereas the second part is the reduction due
to the nuclear potential.  We see that the fusion barrier heights
in all three proximity potentials can be parametrized using the
following relation:
\begin{equation}
V^{par}_{B} = \delta [\frac{1.44Z_1
Z_2}{R_B^{par}}(1-\frac{0.75}{R_B^{par}})]. \label{eq:18}
\end{equation}
Where $\delta$ is a constant having values 0.99903, 0.99868, and
1.002 for Prox 77, Prox 88, and Prox 00, respectively. Here second
term in the above relation is introduced to take care of the
deviations that happen in the lower tail of the fusion barrier
heights. We see that one can parametrize the barrier heights very
closely. The quality of our analytical parametrization is tested
in the Fig. 4, where again percentage difference between
parametrized and exact values are shown. Mathematically,
\begin{equation}
\Delta V_{B}~(\%) = \frac{V_{B}^{par}-
V_{B}^{exact}}{V_{B}^{exact}}\times 100. \label{eq:19}
\end{equation}
Very encouragingly, we see that our fits are within $\pm1\%$ of
the actual values. Some slight deviations can be seen for lighter
masses. This may also be due to the limitations of proximity
potentials in handling the lighter masses where surface is of the
order of nuclear radius.  It is very encouraging to note that our
parametrized form give barrier heights and positions within $\pm
1\%$ of the actual values. The average deviations are -0.10\%,
-0.12\%, and 0.07\% for Prox 77, Prox 88, and Prox 00,
respectively. In Table 1, we display the actual and analytical
parametrized values of some selected collisions for all three
versions of proximity potentials. We note that our results are in
very close agreement with the actual value and therefore,
introduces great simplification in the calculation of fusion
barriers. Finally, we compare our outcome with experimental data
in Fig. 5. Here we display our analytically parametrized
calculated fusion barrier heights $V_{B}^{par}$
[Eq.~(\ref{eq:18})] with experimentally extracted fusion barrier
heights $V_{B}^{expt}$. The experimentally extracted fusion
barrier heights displayed in this figure are obtained in the
approach, when shapes of both colliding nuclei are spherical. The
experimental data are taken from Refs~\cite{ms2000,wang06,expt}.
It is clear from the figure that our results are in good agreement
with experimental data. In a recent attempt~\cite{id}, we
presented comparison of 16 different proximity based potentials
and found that potentials due to Bass~\cite{wr94}, Aage
Winther~\cite{aw95}, and Denisov~\cite{id}~(marked as Bass 80, AW
95, and Denisov DP in Ref.~\cite{id}) were performing better than
other proximity based potentials. The analytical parametrizations
of such potentials will be presented elsewhere~\cite{idd}.
\section{\label{summary}Summary}
Using three versions of proximity potentials, we obtained
analytical relations for the fusion barrier heights and positions.
Our analysis  is based on the calculations of 400 reactions. Our
analytical parametrized values are in very close agreement with
actual as well as experimental values. Therefore, introducing
great simplifications in the calculation of fusion barrier heights
and positions. These results can be used as a guide line for
estimating the fusion barriers in those cases where measurements
do not exist and also for the study
of new nuclei yet unexplored. \\
\section{Acknowledgments}
This work was supported by a research grant from the Department of
Atomic Energy, Government of India.

\newpage
\begin{figure}
\includegraphics* [scale=0.4] {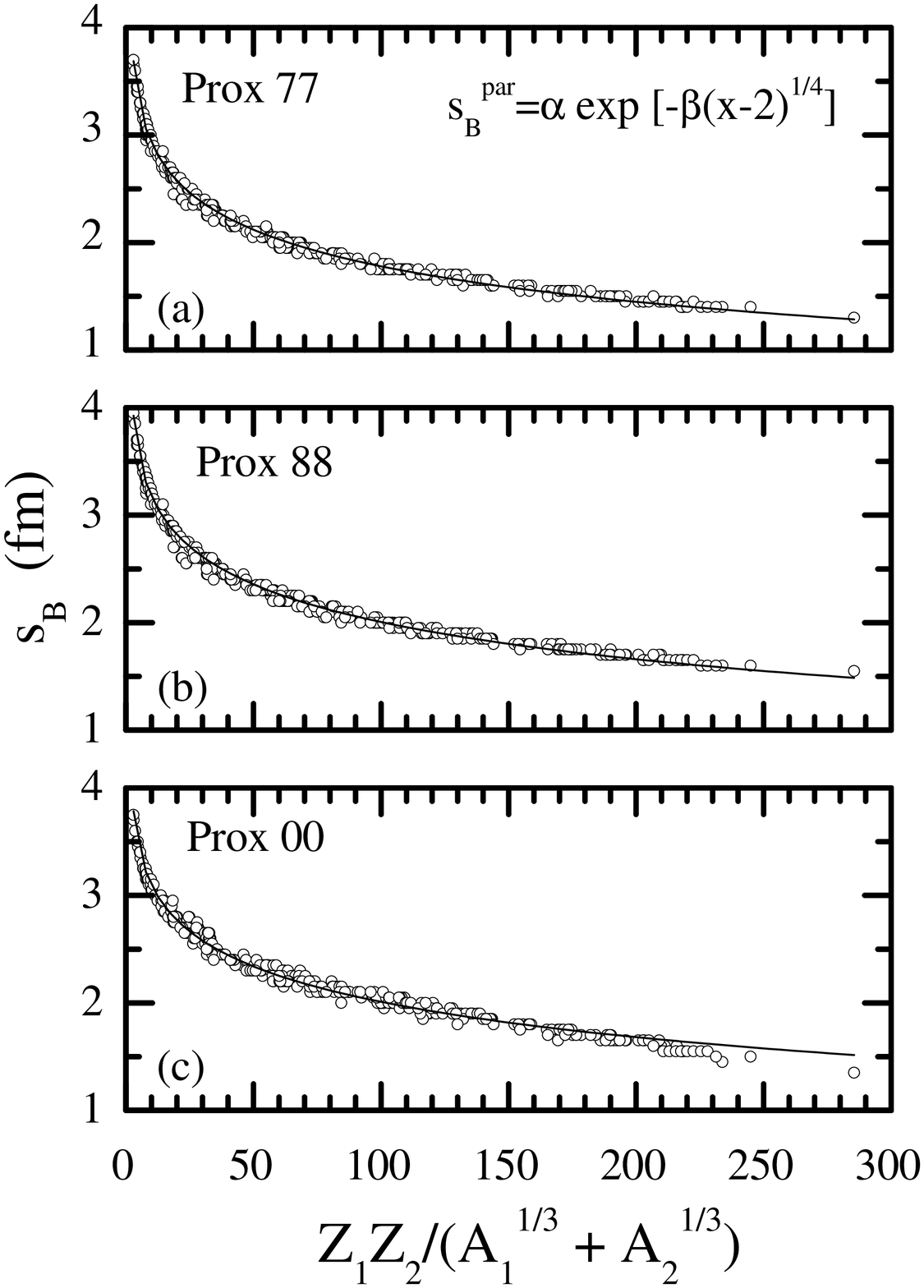}
\vskip -0.2 cm \caption {Reduced  fusion barrier positions
$s_{B}~\rm (fm)$ (defined as
 $s_{B}=R_{B}-C_{1}-C_{2}$) as  a function of the $\frac{Z_1
Z_2}{A_1^{1/3}+A_2^{1/3}}$. Parts (a), (b), and (c) show the
results with Prox 77, Prox 88, and Prox 00 versions of the
proximity potential. Our parametrized fits are shown as solid
curves. The values of constants $\alpha$ and $\beta$ are given in
the text.}\label{fig1} \vskip -0.4 cm
\end{figure}
\begin{figure}
\includegraphics* [scale=0.4] {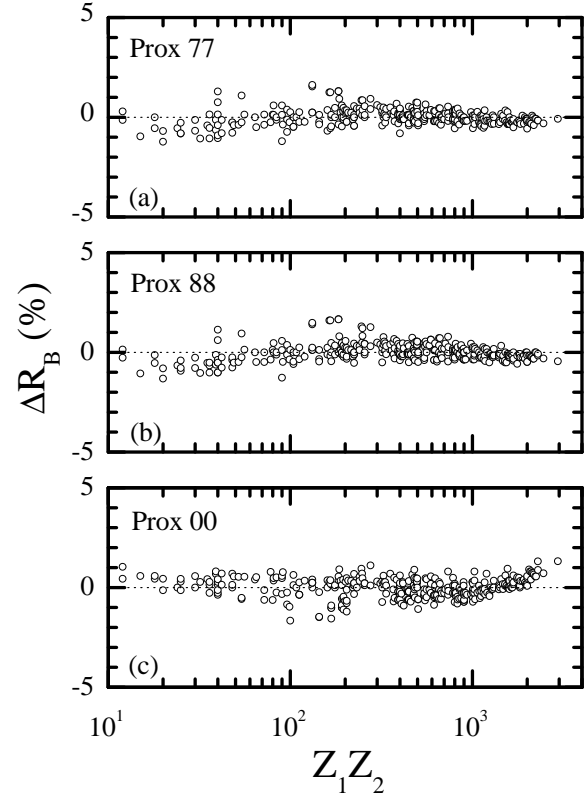}
\vskip -0.2 cm \caption {The  percentage difference $\Delta
R_{B}~(\%)$ [defined in Eq. (\ref{eq:17})] as a function of the
product of charges of colliding pair $Z_{1}Z_{1}$. Parts (a), (b),
and (c) show the results with Prox 77, Prox 88, and Prox 00
versions of the proximity potential.}\label{fig2} \vskip -0.4 cm
\end{figure}
 \begin{figure}
\includegraphics* [scale=0.4] {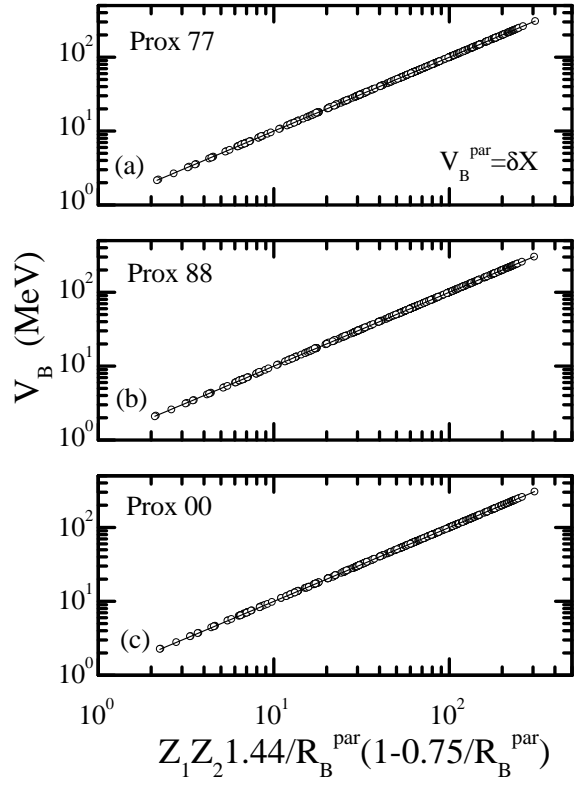}
\vskip -0.2 cm \caption {The fusion barrier heights $V_{B}$ (MeV),
as a function of
 $\frac{1.44Z_1
Z_2}{R_B^{par}}(1-\frac{0.75}{R_B^{par}})$. Parts (a), (b), and
(c) show the results with Prox 77, Prox 88, and Prox 00 versions
of the proximity potential. Our parametrized fits are shown as
solid curves. The value of the constant $\delta$ is given in the
text.}\label{fig3} \vskip -0.4 cm
\end{figure}

\begin{figure}
\includegraphics* [scale=0.4] {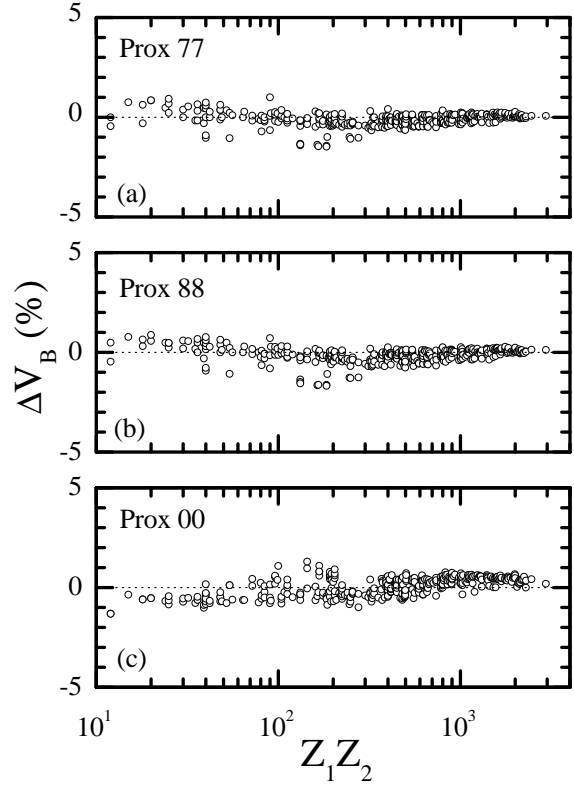}
\vskip -0.2 cm \caption {Same as Fig. 2, but for $\Delta
V_{B}~(\%$).}\label{fig4} \vskip -0.4 cm
\end{figure}
\begin{figure}
\centering
\includegraphics* [scale=0.4] {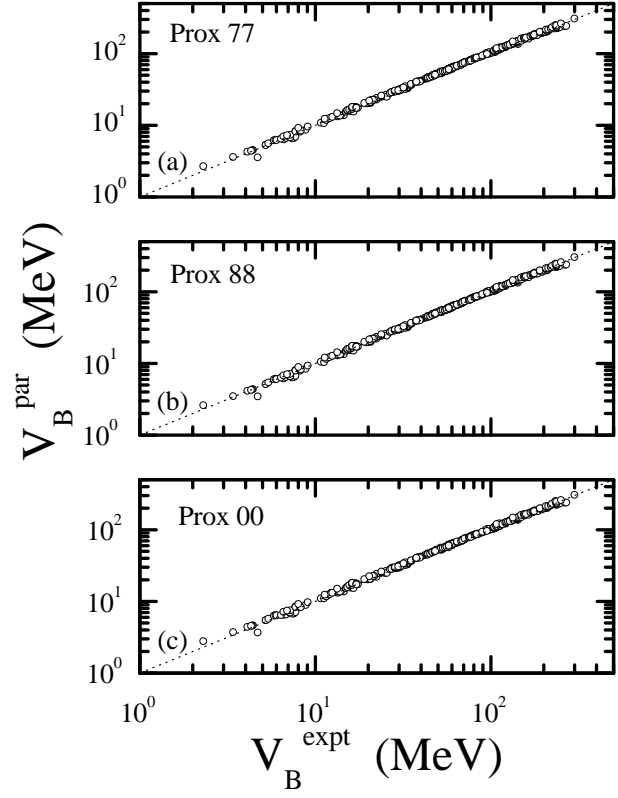}
\vskip -0.2 cm \caption {The variation of the parametrized fusion
barrier heights $V_{B}^{par}~\rm (MeV)$ as a function of
experimental fusion barrier heights $V_{B}^{expt}~\rm (MeV)$.
Parts (a), (b), and (c) show the results with Prox 77, Prox 88,
and Prox 00 versions of the proximity potential. The experimental
values are taken from
Refs.~\cite{ms2000,wang06,expt}.}\label{fig4} \vskip -0.4 cm
\end{figure}
\newpage
\begin{table*}
 \caption{\label{table1}Fusion barrier heights V$_{B}$ (in MeV) and
positions R$_{B}$ (in fm), calculated using different proximity
potentials along with their corresponding parametrized values are
displayed for few cases.}
\begin{tabular}{ccccccccccccccccccccc} \hline \hline
Reaction &\multicolumn{2}{c}{Prox 77}&\multicolumn{2}{c}{Prox
88}&\multicolumn{2}{c}{Prox 00}
&\multicolumn{2}{c}{Prox 77}&\multicolumn{2}{c}{Prox 88}&\multicolumn{2}{c}{Prox 00}\\
\hline &R$_B^{exact}$&R$_B^{par}$ &R$_B^{exact}$&R$_B^{par}$
&R$_B^{exact}$&R$_B^{par}$ &V$_B^{exact}$&V$_B^{par}$
&V$_B^{exact}$&V$_B^{par}$ &V$_B^{exact}$&V$_B^{par}$
\\ \hline
$^{6}$Li + $^{9}$Be      &7.01   &7.03   &7.26   &7.27   &6.74   &6.81    &2.21    &2.20    &2.14    &2.13    &2.29    &2.26      \\
$^{10}$B + $^{12}$C      &7.22   &7.21   &7.47   &7.45   &6.99   &7.03    &5.36    &5.36    &5.19    &5.20    &5.54    &5.50      \\
$^{16}$O + $^{16}$O      &7.65   &7.65   &7.90   &7.90   &7.51   &7.54    &10.86   &10.86   &10.55   &10.55   &11.10   &11.03     \\
$^{20}$Ne + $^{20}$Ne    &7.95   &7.97   &8.20   &8.21   &8.42   &8.28    &16.39   &16.35   &15.94   &15.92   &15.68   &15.85     \\
$^{24}$Mg + $^{26}$Mg    &8.40   &8.37   &8.65   &8.61   &8.86   &8.73    &22.54   &22.53   &21.95   &21.96   &21.47   &21.75     \\
$^{24}$Mg + $^{34}$S     &8.61   &8.61   &8.86   &8.85   &8.89   &8.80    &29.34   &29.28   &28.60   &28.55   &28.64   &28.80        \\
$^{16}$O + $^{64}$Ni     &9.01   &9.03   &9.26   &9.27   &9.05   &9.08    &35.17   &35.06   &34.33   &34.22   &35.08   &34.99        \\
$^{6}$Li + $^{238}$U     &10.87  &10.97  &11.07  &11.21  &10.81  &10.93   &34.07   &33.72   &33.46   &33.04   &34.28   &33.94     \\
$^{12}$C + $^{124}$Sn    &9.88   &9.94   &10.13  &10.18  &9.97   &10.00   &40.31   &40.14   &39.49   &39.26   &40.20   &40.04      \\
$^{16}$O + $^{110}$Pd    &9.88   &9.90   &10.08  &10.13  &10.02  &10.01   &49.60   &49.42   &48.56   &48.38   &49.12   &49.07    \\
$^{30}$Si + $^{64}$Ni    &9.63   &9.60   &9.83   &9.84   &9.71   &9.65    &54.13   &54.16   &52.94   &52.92   &53.93   &54.06     \\
$^{48}$Ca + $^{48}$Ca    &9.89   &9.81   &10.09  &10.05  &9.89   &9.83    &53.96   &54.18   &52.84   &52.97   &53.93   &54.24     \\
$^{32}$S + $^{58}$Ni     &9.40   &9.45   &9.65   &9.68   &9.50   &9.53    &63.04   &62.79   &61.60   &61.40   &62.64   &62.49     \\
$^{40}$Ar + $^{60}$Ni    &9.82   &9.78   &10.02  &10.02  &10.00  &9.94    &68.40   &68.45   &66.91   &66.92   &67.37   &67.64      \\
$^{16}$O + $^{166}$Er    &10.64  &10.66  &10.84  &10.89  &10.77  &10.76   &68.56   &68.25   &67.25   &66.89   &67.93   &67.87      \\
$^{16}$O + $^{186}$W     &10.86  &10.90  &11.06  &11.13  &11.18  &11.15   &73.09   &72.76   &71.74   &71.34   &71.39   &71.45      \\
$^{36}$S + $^{90}$Zr     &10.30  &10.28  &10.55  &10.50  &10.41  &10.36   &82.99   &83.03   &81.30   &81.39   &82.35   &82.69     \\
$^{35}$Cl + $^{92}$Zr    &10.25  &10.25  &10.50  &10.47  &10.39  &10.36   &88.58   &88.45   &86.75   &86.71   &87.64   &87.85     \\
$^{32}$S + $^{110}$Pd    &10.43  &10.45  &10.68  &10.68  &10.65  &10.65   &94.21   &94.05   &92.33   &92.15   &92.43   &92.70      \\
$^{64}$Ni + $^{64}$Ni    &10.48  &10.47  &10.73  &10.70  &10.60  &10.57   &99.84   &100.00  &97.86   &97.98   &98.90   &99.43     \\
$^{40}$Ar + $^{110}$Pd   &10.75  &10.73  &10.95  &10.95  &11.07  &10.98   &103.19  &103.25  &101.21  &101.30  &100.61  &101.37      \\
$^{32}$S + $^{138}$Ba    &10.87  &10.87  &11.07  &11.09  &10.93  &10.96   &110.71  &110.40  &108.62  &108.33  &109.73  &109.89    \\
$^{40}$Ar + $^{130}$Te   &11.05  &11.03  &11.25  &11.26  &11.22  &11.18   &113.63  &113.78  &111.56  &111.58  &111.96  &112.69     \\
$^{24}$Mg + $^{208}$Pb   &11.41  &11.44  &11.61  &11.66  &11.73  &11.69   &116.04  &115.63  &114.02  &113.56  &113.09  &113.66     \\
$^{29}$Si + $^{178}$Hf   &11.27  &11.28  &11.47  &11.50  &11.55  &11.49   &120.24  &120.00  &118.08  &117.83  &117.75  &118.32      \\
$^{34}$S + $^{168}$Er    &11.35  &11.32  &11.55  &11.55  &11.39  &11.40   &129.16  &129.10  &126.86  &126.67  &128.04  &128.65     \\
$^{64}$Ni + $^{96}$Zr    &11.13  &11.08  &11.33  &11.30  &11.21  &11.19   &135.37  &135.58  &132.87  &133.07  &134.04  &134.74    \\
$^{38}$S + $^{181}$Ta    &11.69  &11.64  &11.89  &11.87  &11.79  &11.78   &134.80  &135.05  &132.51  &132.56  &133.21  &133.96       \\
$^{48}$Ca + $^{154}$Sm   &11.61  &11.59  &11.86  &11.80  &11.72  &11.68   &143.72  &143.95  &141.26  &141.51  &142.55  &143.35     \\
$^{40}$Ar + $^{180}$Hf   &11.65  &11.66  &11.90  &11.88  &11.81  &11.80   &149.63  &149.61  &147.07  &146.98  &147.58  &148.40     \\
$^{38}$S + $^{208}$Pb    &11.98  &11.94  &12.18  &12.16  &12.00  &12.00   &147.89  &148.15  &145.47  &145.60  &147.31  &147.90     \\
$^{64}$Ni + $^{124}$Sn   &11.55  &11.52  &11.75  &11.73  &11.68  &11.68   &163.23  &163.45  &160.37  &160.67  &160.85  &161.84    \\
$^{40}$Ar + $^{206}$Pb   &11.93  &11.94  &12.18  &12.16  &12.11  &12.10   &166.66  &166.67  &163.89  &163.79  &164.19  &165.10     \\
$^{86}$Kr + $^{100}$Mo   &11.59  &11.57  &11.84  &11.79  &11.68  &11.70   &175.40  &175.81  &172.33  &172.69  &173.67  &174.51    \\
$^{90}$Zr + $^{90}$Zr    &11.42  &11.42  &11.67  &11.64  &11.56  &11.59   &188.23  &188.32  &184.79  &184.94  &185.53  &186.30     \\
$^{40}$Ar + $^{238}$U    &12.31  &12.28  &12.51  &12.49  &12.30  &12.35   &182.29  &182.15  &179.41  &179.22  &181.07  &181.72    \\
$^{96}$Mo + $^{100}$Mo   &11.75  &11.72  &11.95  &11.93  &11.81  &11.86   &202.39  &202.67  &198.85  &199.28  &200.05  &201.03     \\
$^{54}$Cr + $^{196}$Os   &12.22  &12.19  &12.42  &12.40  &12.34  &12.34   &201.86  &202.01  &198.62  &198.75  &199.21  &200.31     \\
$^{51}$V + $^{208}$Pb    &12.23  &12.24  &12.48  &12.45  &12.36  &12.40   &208.11  &208.09  &204.75  &204.73  &205.18  &206.18     \\
$^{54}$Cr + $^{209}$Bi   &12.33  &12.32  &12.53  &12.53  &12.59  &12.61   &218.37  &218.45  &214.85  &214.95  &212.95  &214.38    \\
$^{96}$Zr + $^{124}$Sn   &12.15  &12.13  &12.40  &12.34  &12.28  &12.29   &222.18  &222.53  &218.53  &218.91  &219.15  &220.48     \\
$^{55}$Mn + $^{208}$Pb   &12.35  &12.32  &12.55  &12.53  &12.24  &12.35   &224.74  &224.80  &221.13  &221.20  &224.89  &224.96      \\
$^{70}$Zn + $^{176}$Yb   &12.35  &12.31  &12.55  &12.52  &12.36  &12.41   &230.12  &230.47  &226.42  &226.76  &228.67  &229.41   \\
$^{58}$Fe + $^{208}$Pb   &12.39  &12.40  &12.64  &12.61  &12.38  &12.47   &232.38  &232.38  &228.67  &228.68  &231.26  &231.85     \\
$^{59}$Co + $^{208}$Pb   &12.42  &12.41  &12.62  &12.62  &12.50  &12.57   &241.20  &241.15  &237.34  &237.30  &237.99  &238.98     \\
$^{59}$Co + $^{209}$Bi   &12.43  &12.42  &12.63  &12.63  &12.62  &12.69   &244.02  &243.90  &240.10  &240.01  &238.47  &239.75    \\
$^{63}$Cu + $^{197}$Au   &12.39  &12.37  &12.59  &12.57  &12.20  &12.36   &250.40  &250.29  &246.33  &246.46  &251.22  &251.22     \\
$^{64}$Ni + $^{208}$Pb   &12.56  &12.54  &12.76  &12.75  &12.53  &12.64   &247.56  &247.65  &243.66  &243.74  &245.68  &246.54    \\
$^{70}$Zn + $^{208}$Pb   &12.71  &12.67  &12.91  &12.87  &12.76  &12.85   &262.60  &262.78  &258.53  &258.86  &259.01  &260.10    \\
$^{86}$Kr + $^{208}$Pb   &12.99  &12.98  &13.24  &13.18  &12.92  &13.09   &308.05  &308.27  &303.40  &303.77  &306.16  &306.75    \\
\hline \hline
\end{tabular}
\end{table*}
\end{document}